\documentclass[fleqn,twoside]{article}
 \usepackage{espcrc2}

\usepackage{amsmath}
\usepackage{amssymb}
\usepackage{amsfonts}

\readRCS
$Id: espcrc2.tex,v 1.2 2004/02/24 11:22:11 spepping Exp $
\usepackage{graphicx}
\usepackage[figuresright]{rotating}

\title{New precise determination of the $\tau$ lepton mass at KEDR
  detector\thanks{
    Partially supported by the Russian Foundation for Basic
    Research, Grants 01-02-17477, 04-02-16745-a, 05-02-16594-a,
    05-02-16798-a and the Presidential Grant for Scientific
    School Support NSh-905.2006.2.}
}

\author{
KEDR collaboration\\*[0.75ex]
V.V.~Anashin,
V.M.~Aulchenko,
E.M.~Baldin,
A.K.~Barladyan,
A.Yu.~Barnyakov, 
M.Yu.~Barnyakov, 
S.E.~Baru, 
I.V.~Bedny,
O.L.~Beloborodova,
A.E.~Blinov,
V.E.~Blinov,
A.B.~Bobrov,
V.S.~Bobrovnikov,
A.V.~Bogomyagkov, 
A.E.~Bondar, 
D.V.~Bondarev,
A.R.~Buzykaev,
V.P.~Cherepanov,  
S.I.~Eidelman,
Yu.M.~Glukhovchenko,  
V.V.Gulevich,
S.E.~Karnaev,
G.V.~Karpov, 
S.V.~Karpov,
V.A.~Kiselev,
S.A.~Kononov,
K.Yu.~Kotov,
E.A.~Kravchenko,
E.V.~Kremyanskaya,
V.F.~Kulikov,
G.Ya.~Kurkin,
E.A.~Kuper,
E.B.~Levichev,
D.A.~Maksimov,
V.M.~Malyshev,
A.L.~Maslennikov,
A.S.~Medvedko,
O.I.~Meshkov,
S.E.~Mishnev,
I.I.~Morozov,
N.Yu.~Muchnoi,
V.V.~Neufeld,
S.A.~Nikitin,
I.B.~Nikolaev,
A.P.~Onuchin,
S.B.~Oreshkin,
I.O.~Orlov,
A.A.~Osipov,
S.V.~Peleganchuk,
S.S.~Petrosyan,
S.G.~Pivovarov,
P.A.~Piminov,
V.V.~Petrov,
A.O.~Poluektov,
G.E.~Pospelov, 
V.G.~Prisekin,
A.A.~Ruban,
V.K.~Sandyrev, 
G.A.~Savinov,
A.G.~Shamov,
D.N.~Shatilov,
E.I.~Shubin, 
B.A.~Shwartz,
V.A.~Sidorov,
E.A.~Simonov,
S.V.~Sinyatkin, 
Yu.I.~Skovpen,
A.N.~Skrinsky,
V.V.~Smaluk,
A.M.~Soukharev,
M.V.~Struchalin,
A.A.~Talyshev,
V.A.~Tayursky,
V.I.~Telnov,
Yu.A.~Tikhonov,
K.Yu.~Todyshev,
G.M.~Tumaikin,
Yu.V.~Usov,
A.I.~Vorobiov,
A.N.~Yushkov,
V.N.~Zhilich,
A.N.~Zhuravlev\\*[0.75ex]
presented by B.A.~Shwartz
\address[BINP]{Budker Institute of Nuclear Physics,
  630090 Novosibirsk, Russia}
}

\runtitle{New precise determination of the $\tau$ lepton mass at KEDR detector}
\runauthor{KEDR Collaboration}

\begin{document}

\begin{abstract}
The status of the experiment on the precise $\tau$ lepton mass 
measurement running at the VEPP-4M
collider with the KEDR detector is reported. The mass value is evaluated
from the $\tau^+\tau^-$ cross section behaviour around the production
threshold. The preliminary result based on 6.7 pb$^{-1}$ of data is
    \mbox{$m_{\tau} = 1776.80^{+0.25}_{-0.23} \pm 0.15$}~MeV.
Using 0.8 pb$^{-1}$ of data collected at the $\psi'$ peak the preliminary
result is also obtained:
\mbox{$\Gamma_{ee}\!\cdot\!B_{\tau\tau} (\psi') = 7.2 \pm 2.1$}~eV.
\vspace*{1pc}
\end{abstract}

\maketitle

\section{Introduction}

The $\tau$ lepton mass, $m_\tau$, is one of the  fundamental 
characteristics of the Standard Model. Together with the lifetime and
the decay probability to $e\bar{\nu}_{e}\nu_{\tau}$ this value can be used
to test the lepton universality principle which is one
of the postulates of the modern Electroweak theory.
The world average value \mbox{$m_\tau=1776.99^{+0.29}_{-0.26}$}
\cite{PDG2006} is
dominated by the result of the BES collaboration \cite{BESfinal}
which statistical analysis
and uncertainty estimations were recently discussed in Refs.~\cite{TAU04}
and \cite{BEScomment}.
Thus, the additional measurements are desirable to improve 
the mass accuracy and ensure
future progress in the lepton universality tests.

The direct method of the $\tau$ mass determination is
a study of the threshold behaviour of the  $\tau^{+}\tau^{-}$
production cross section in $e^{+}e^{-}$ collisions
as it was done in the experiments ~\cite{DELCO} and then \cite{BESfinal}.
The key  question of such experiments is the precision in
the beam energy determination. The important feature of the present 
work is an application of two independent methods of the beam energy 
measurement, while the previous experiments relied on the extrapolation
based on the $J/\psi$ and $\psi'$ mesons as reference points.  
It should be noted as well that the beam energy in our experiment
is monitored with the accuracy better than $5\cdot10^{-5}$ and the
absolute energy calibration is done with the precision of
$1\cdot10^{-5}$. 

\section{\label{sec:VEPP}VEPP-4M collider and KEDR detector}

\begin{figure}[t]
\centering
\vspace*{-2.2cm}
\hspace*{-3.1cm}
\includegraphics[clip,height=0.57\textwidth]{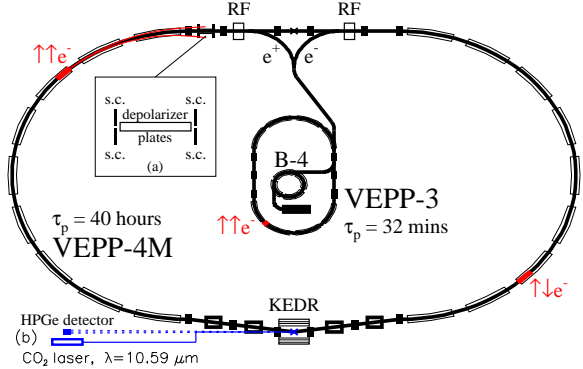}
\vspace*{-3.2cm}
\caption{\label{fig:VEPP}VEPP-4M/VEPP-3 accelerator complex in
the energy calibration
mode: (a) -- Touschek polarimeter, (b) -- Compton backscattering monitor;
spin polarization time $\tau_p$ is for 1.85 GeV.
}
\vspace*{-0.7cm}
\end{figure}

The VEPP-4M/VEPP-3 accelerator complex is presented schematically
in Figure~\ref{fig:VEPP}.
The \mbox{VEPP-4M} collider \cite{VEPP4} has the circumference of 366~m and
operates in 2$\times$2 bunches mode. The beam energy can vary
in the range of 1$\div$6 GeV, the peak luminosity at the $\tau$--production
threshold $E_{beam} \approx 1.78$ GeV is about
$2\!\cdot\!10^{30}$~cm$^{-2}$s$^{-1}$.

 The beams, optionally polarized,
are injected from the \mbox{VEPP-3} booster at the energy
up to 1.9~GeV. This allows to apply the resonant depolarization
method (RDM) \cite{REDE} for the precise energy calibration.
The Touschek (intra-beam scattering) polarimeter of VEPP-4M
(Figure~\ref{fig:VEPP}a) requires special runs for the calibration.
During data taking, the beam energy can be monitored
using the Compton backscattering (CBS) of the infra-red laser light
(Figure~\ref{fig:VEPP}b) by the method developed at the synchrotron light source
BESSY-I \cite{BESSY1}. The statistical accuracy of the single
measurement is about 100~keV, the systematic uncertainty
of the method verified by the resonant depolarization is close
to 60~keV.

The KEDR detector \cite{KEDR} consists of the vertex detector, the
drift chamber, the time-of-flight system of scintillation counters,
the particle identification system based on the aerogel Cherenkov
counters, the calorimeter with the longitudinal segmentation 
(the liquid krypton in the barrel part and
the CsI crystals in the end caps) and the muon tube system inside 
the magnet yoke. Currently KEDR operates at the magnetic
field of 6~kGs.

The longitudinal segmentation of the calorimeter provides good
$e/\pi$ identification used to select $\tau^{+}\tau^{-}$ events.

\section{\label{seq:SCAN}Experiment scenario}

\begin{figure}[t]
\centering
\vspace*{-0.6cm}
\hspace*{-0.6cm}
\includegraphics[clip,width=0.55\textwidth]{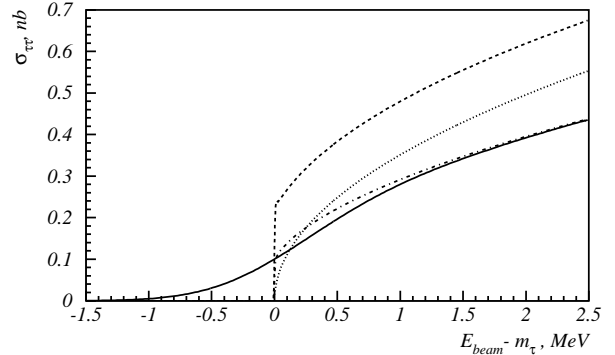}
\vspace*{-1.6cm}
\caption{\label{fig:XS}$e^+e^- \to \tau^+\tau^-$ cross section near
threshold  as function of the beam energy (dotted line -- Born approximation;
dashed line -- plus the Coulomb interaction, the final state radiation
and the vacuum polarization;
dash-dotted line -- plus initial state radiation;
solid line -- plus the beam energy spread).
}
\vspace*{-0.7cm}
\end{figure}

A cross section of the process $e^+e^- \to \tau^+\tau^-$ measured 
at certain center-of-mass energy $W$ is expressed as 
\begin{equation}\label{eq:SigmaTT}
\begin{split}
 \sigma(W) = & \frac{1}{\sqrt{2\pi}\sigma_W} \!\int dW'\!
\exp\left\{-\frac{(W\!-\!W')^2}{2\sigma_W^2} \right\} \\
& \int\! dx\, F(x,W') \,\sigma_{fs}(W'\sqrt{1-x}),
\end{split}
\end{equation}
\vspace*{-0.325cm}

\noindent
where the first integral stands to take into account c.m.s.
energy spread, $\sigma_W$, the second one accounts the energy
loss due to the initial state radiation \cite{KF}, while
\begin{equation}\label{eq:SigmaFS}
\sigma_{fs}(W)=\frac{4\pi\alpha^2}{3W^2}\frac{\beta(3-\beta^2)}{2}
\frac{F_c(\beta)F_r(\beta)}{|1\!-\!\Pi(W)|^2}
\end{equation}
includes 
the Coulomb interaction correction
\mbox{$F_c(\beta)\!=\!(\pi\alpha/\beta)/(1\!-\!\exp{(-\pi\alpha/\beta)})$},
the final state radiative correction $F_r(\beta)$ \cite{VOLOSHIN} and
the vacuum polarization effect $|1\!-\!\Pi(W)|^2$.
The quantity \mbox{$\beta\!=\!(1\!-\!(2m_{\tau}/W)^2)^{1/2}$}
is the $\tau$ lepton velocity.

Due to Coulomb interaction of the produced $\tau^{+}$ and $\tau^{-}$ 
the cross section~(\ref{eq:SigmaFS}) energy dependence has a step
at $W\!=\!2m_\tau$ (Figure\ref{fig:XS}).

The narrow region
of a few MeV around the threshold is most sensitive to the mass
value. For this reason the following scan scenario was chosen:
70\% of the integrated luminosity $\mathcal{L}$ are taken at three points
$E_{beam}\!\!=m_\tau\!-\!0.5,\: m_\tau,\: m_\tau\!+\!0.5$~MeV with
the world average value of $m_\tau$, 15\% of the data
are collected well below the
threshold to fix the background level $\sigma_B$ and remaining 15\% --
well above the threshold to determine the effective detection
efficiency $\varepsilon$.
The interval of $\,\pm$0.5~MeV covers possible  uncertainty of the mass;
a few additional points above the threshold were foreseen to increase the
robustness of the three-parameter data fit.

\section{Beam energy determination}

\begin{figure}[t]
\centering
\vspace*{-0.5cm}
\hspace*{-0.25cm}
\includegraphics[clip,width=0.5\textwidth]{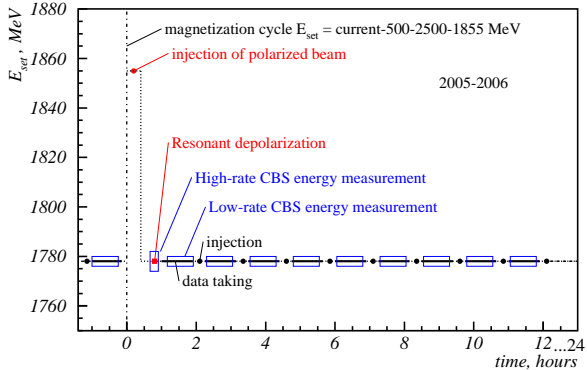}
\vspace*{-1.4cm}
\caption{\label{fig:VEPP-SC}The VEPP-4M operation scenario in 2005-2006 (in
2004-2005 only high-rate Compton backscattering measurements were used,
incompatible with the data taking). }
\vspace*{-0.7cm}
\end{figure}

A conventional way of the beam energy determination is 
a calculation based on the measured magnet currents. It
provides the relative accuracy that seems to be
not better than $3\!\cdot\!10^{-4}$.
The uncontrollable energy variations are of the same order of magnitude.
Thus the precise beam energy calibration
is required for the $\tau$ mass determination
and, at least, the reliable energy stability
tests are necessary for an accurate uncertainty estimate. 

\begin{figure}[t]
\centering
\hspace*{-0.3cm}
\includegraphics[width=0.475\textwidth,clip,viewport=0 0 510 349]
{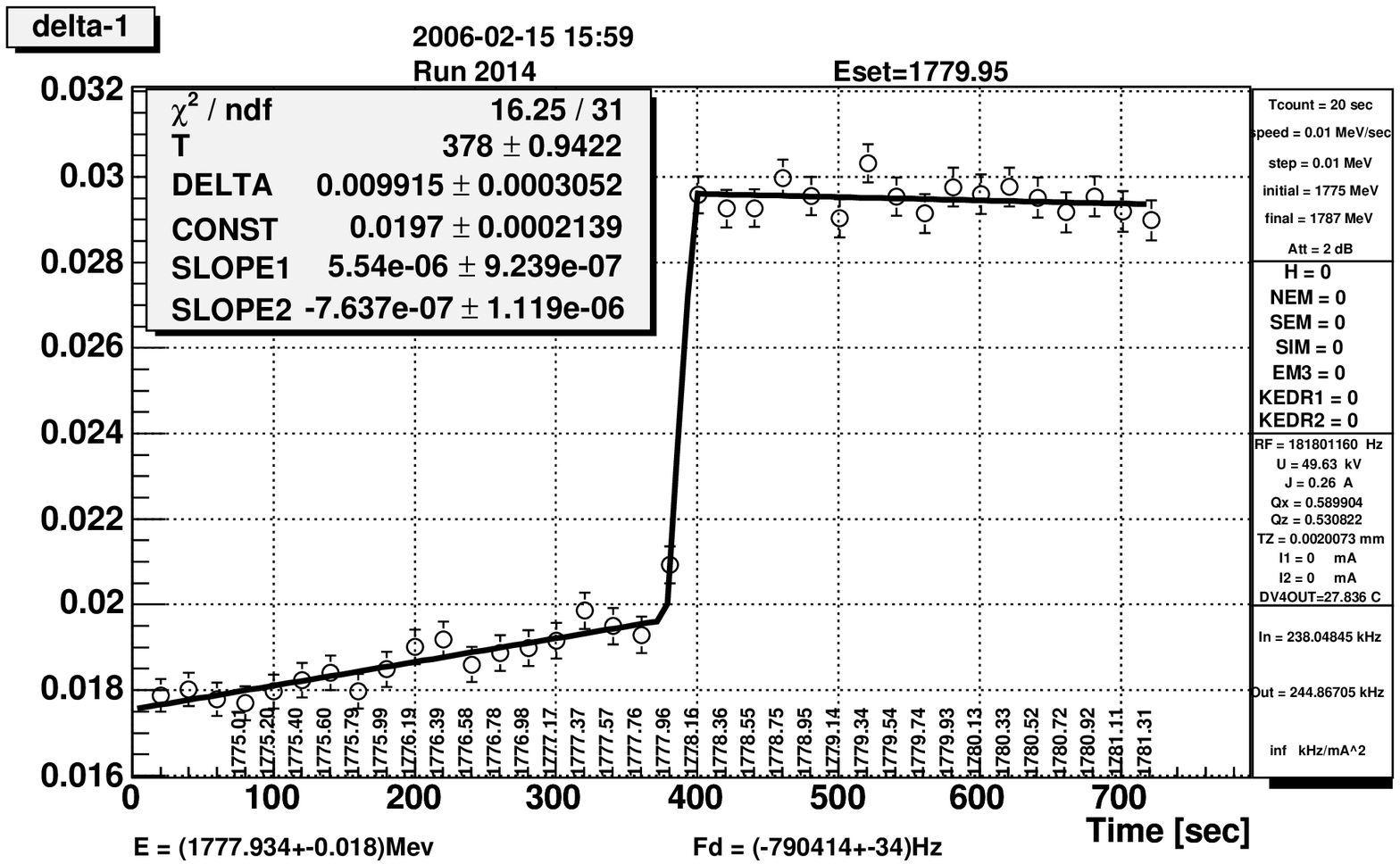}
\vspace*{-1.4cm}
\caption{\label{fig:Jump}A typical resonant depolarization run: the ratio
of the intrabeam scattering rates from the unpolarized and polarized
bunches minus one.}
\vspace*{-0.7cm}
\end{figure}

In the previous KEDR experiments on the high precision $J/\psi$ and
$\psi'$ meson mass measurements \cite{PSI2002} various sources of the systematic
uncertainties in the beam energy determination were thoroughly studied
to achieve a 10~keV accuracy.

In this experiment the basic energy calibrations were performed
by the resonant depolarization with the smoothing interpolation of the RDM results
between the calibrations as described in Ref.~\cite{PSI2002}
(the guiding field measurements and the ring and the tunnel temperature measurements
are employed for the interpolation).

The improvements of the Touschek
polarimeter (Figure~1a) done since 2003 have allowed to operate at
$E_{beam}\!\approx\!1772$~MeV, where the polarization
lifetime is $\lesssim\!1000$ sec because of the closeness of the
integer spin resonance $\nu\!=\!4$ (1762.59~MeV). However, the absence
of polarization in \mbox{VEPP-3} at the energy region of
$1700\!\div\!1830$~MeV forced to employ the complicated
machine operation scenario shown in Figure~\ref{fig:VEPP-SC}.
After staying in the threshold region
the magnetization cycle must be performed in \mbox{VEPP-4M} to inject the
polarized beam above the region quoted.
This and also some forced changes in the accelerator cooling system reduced
the accuracy of the energy interpolation
between the calibrations from 8~keV obtained in ~\cite{PSI2002}
to 30~keV.

\begin{figure}[t]
\centering
\hspace*{-0.3cm}
\includegraphics[width=0.48\textwidth,clip,viewport=0 0 567 280]{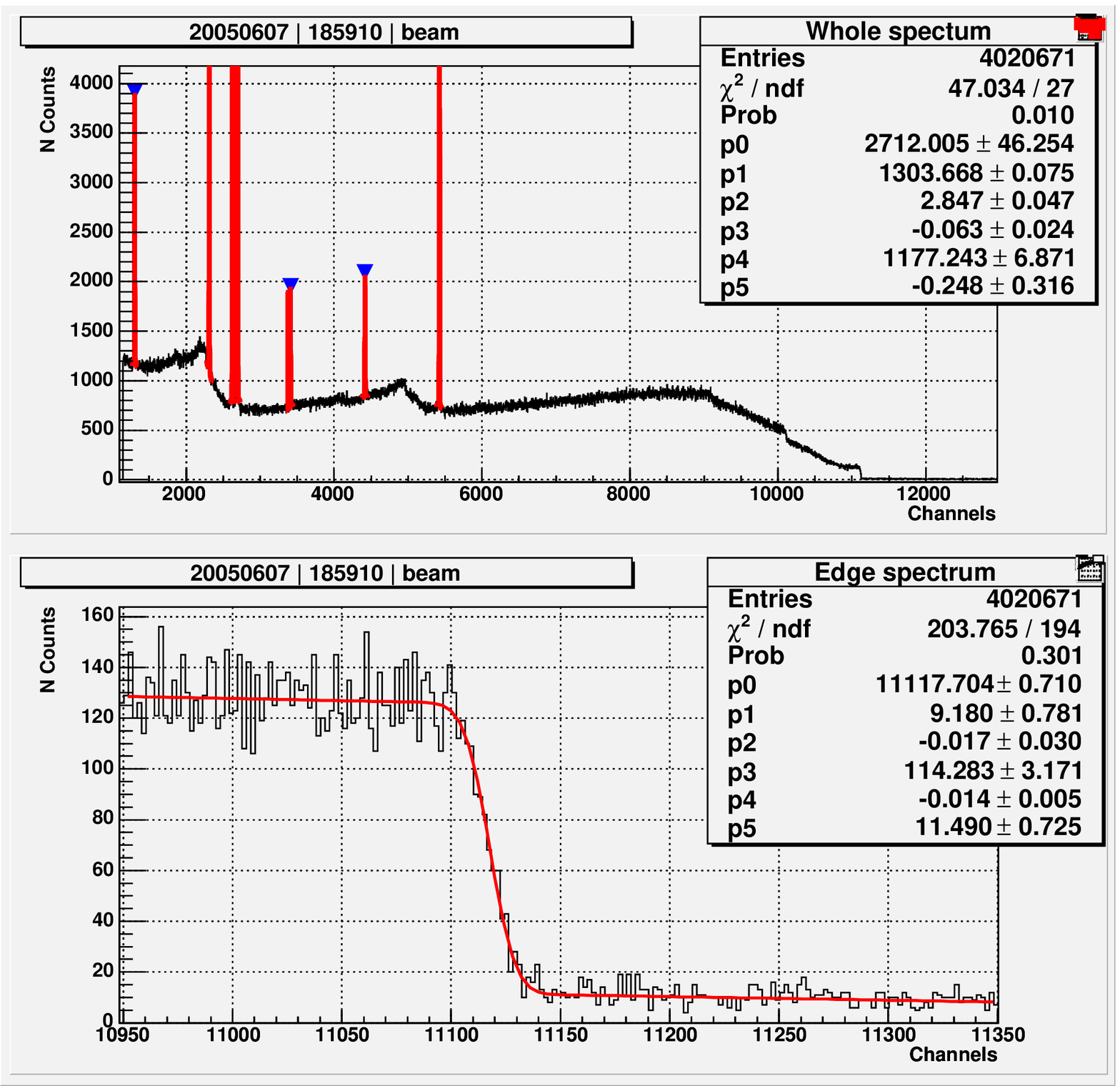}
\vspace*{-1.1cm}
\caption{\label{fig:CBS}A typical fit of the Compton backscattering spectrum edge
(5.78 MeV) accounting for the background and the detection efficiency variations.}
\vspace*{-0.5cm}
\end{figure}

The resonant depolarizations were performed normally once
per day
with the accuracy better than 20~keV.
The results of the typical resonant depolarization run is shown
in Figure~\ref{fig:Jump}.
Between the depolarizations the energy was directly measured 
using the CBS monitor (Sec.~\ref{sec:VEPP} and Figure~\ref{fig:VEPP-SC})
with the statistical accuracy of
about 100~keV. The multiparameter fit of the Compton spectrum edge is shown
in Figure~\ref{fig:CBS}. It accounts for the nonuniform background and the
detection efficiency variations.


\begin{figure}[hb]
\centering
\vspace*{-0.6cm}
\includegraphics[clip,height=0.325\textwidth]{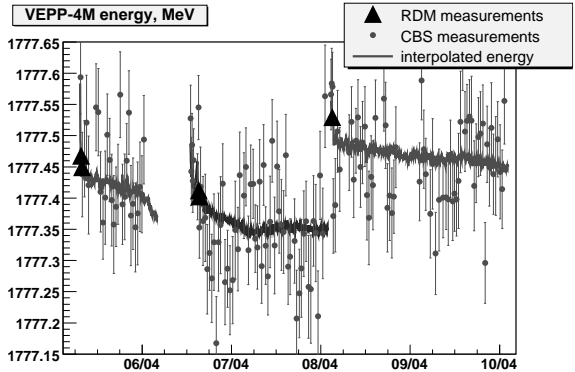}
\vspace*{-1.6cm}
\caption{\label{fig:VEPP-E}An example of VEPP-4M energy behavior,
April 2006.}
\vspace*{-0.7cm}
\end{figure}

An example of VEPP-4M energy behavior during three successive runs 
is presented in Figure~\ref{fig:VEPP-E}.
The RDM measurements were performed at the start of each run.
During the run the energy value were measured by CBS and evaluated
using interpolation. The values obtained by these two methods are
mostly in agreement.  
The systematic difference in the beam energy obtained by the interpolation
of resonant depolarization data and by the CBS measurement sometimes
reaches 100~keV. It's a subject of further investigations;
some corrections can be applied post factum.
The magnetization cycles allow to reproduce the machine
energy with the accuracy $\sim 1\!\cdot\!10^{-4}$, however,
it is not a limiting factor for the mass measurement accuracy.

\section{Energy spread determination}

To calculate the $\tau^{+}\tau^{-}$ cross section from Eq.~\ref{eq:SigmaTT},
the c.m. energy spread $\sigma_W$ must be known with the high
accuracy. The \mbox{VEPP-4M} settings related to the beam energy spread
were optimized for the $\tau$ mass experiment and kept unchanged
since 2004.

The three scans of $\psi'$ and one scan of $J/\psi$ performed
in 2004-2006 to determine $\sigma_W$ in the vicinity of the $\tau$ threshold
resulted in
\vspace*{-0.1cm}
\begin{eqnarray}
\sigma_W(\psi') & \!=\! &1.15 \pm 0.02 \pm 0.03 \text{ MeV},\nonumber\\
\sigma_W(J/\psi) & \!=\! &0.72 \pm 0.01 \pm 0.02 \text{ MeV}.\nonumber
\end{eqnarray}
At the $J/\psi$ peak the 11\% deviation from the expected value
of $\sigma_W(\psi')\!\times\!(M_{J/\psi}/M_{\psi'})^2$ exists.
A similar deviation took place 
during the  $J/\psi$-- and of $\psi'$--mass measurements 
\cite{PSI2002} with the different spread-related settings.

 Assuming the linear growth of the deviation with
\mbox{$W\!-\!M_{\psi^\prime}$} we obtained
\vspace*{-0.15cm}
$$\sigma_W(2m_\tau) = 1.07 \pm 0.02 \pm 0.04 \text{ MeV.}$$
No essential dependence of the energy spread on the beam current
was observed at the $\psi'$ region neither in the resonance scans
nor by means of the beam diagnostic \cite{RUPAC}.

\section{Selection of $\tau$ events}

To diminish systematic uncertainties the event selection criteria
were chosen as loose as possible when a background was kept to
be negligible.
The two-prong events due to
\vspace*{-0.1cm}
\begin{equation}
\begin{split}
\hspace*{-6mm}
 e^+e^- \!\rightarrow\:\, &
 (\tau\rightarrow e\nu_\tau\bar{\nu}_e,\:\mu\nu_\tau\bar{\nu}_\mu,
   \,\pi\nu_\tau,\,K\nu_\tau,\,\rho\nu_\tau) \\ 
 & (\tau\rightarrow e\nu_\tau\bar{\nu}_e)^{*} \\
&  +\text{c.c.}
\end{split}
\nonumber
\end{equation}

\vspace*{-0.15cm}
\noindent
were selected. At least one track must be identified as an electron
using the signal
in the calorimeter and the momentum measurements.
The $\mu$/$\pi$/$K$ identification was not applied;
it does not reduce the systematic uncertainty of the mass.
No photons with $E_{\gamma}\!>\!30$~MeV were allowed.
The other cuts were
$E\!<\!2200$ MeV, $p_T\!>\!200$~MeV, $p_T/(W\!-\!E)\!>\!0.06$, where
$p_T$ is the total transverse momentum,
$E$ is total energy of the detected particles and $W=2E_{beam}$.

\begin{figure}[ht]
\centering
\vspace*{-0.7cm}
\hspace*{-0.65cm}
\includegraphics[width=0.51\textwidth]{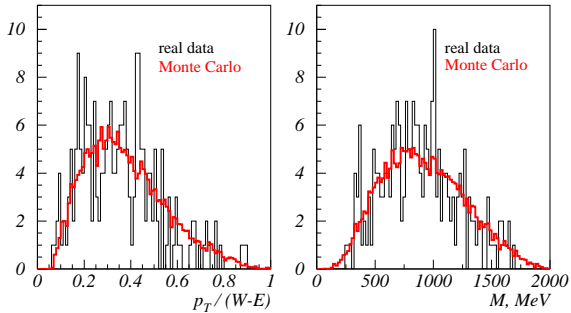}
\vspace*{-1.4cm}
\caption{\label{fig:Hist}The distributions in the $p_T$ over the missing
energy $(W\!\!-E)$ (left) and in the invariant mass of the detected
system (right); the real data (small statistics) and the simulation
(high statistics).
}
\vspace*{-0.7cm}
\end{figure}

With such cuts the residual background (mainly two-photon)
is expected to be uniform in the energy region of the experiment.
According to the Monte Carlo
calculations, the detection efficiency at the
$\tau$ threshold is about 2.5\% with the relative reduction by 
10\% at \mbox{$W=3777$~MeV}.

The distributions in some parameters of interest
for the real data and the simulation
are presented in Figure~\ref{fig:Hist}.

\section{Preliminary results}

\begin{figure}[hb]
\centering
\vspace*{-1.3cm}
\hspace*{-0.3cm}
\includegraphics[width=0.52\textwidth]{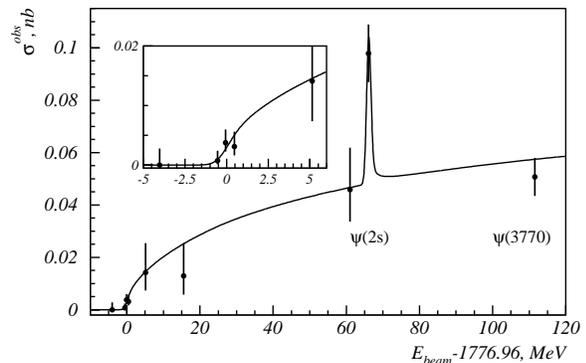}
\vspace*{-1.7cm}
\caption{\label{fig:Scan}The observed $\tau^{+}\tau^{-}$ cross section
versus the beam energy.
}
\vspace*{-0.7cm}
\end{figure}

The preliminary results of the $\tau^{+}\tau^{-}$ threshold scan
are collected in Table~\ref{tab:Points} and presented in Figure~\ref{fig:Scan}.
The energy $\left<E\right>$ assigned to the point is the average of all
measured values.
The corresponding standard deviation
$\delta_E$
is related to the machine energy instability and is much less than
the beam energy spread $\sigma_E \approx \sigma_W/\sqrt{2}$.

To determine the value of $\tau$ lepton mass the log--likelihood fit
of the observed number of events in the nine points were done. The expected
number of events in the point was parameterized as
\begin{equation}
  n_i = 
\left(\varepsilon\, r_i \,\sigma(2\!\left<E\right>_i,m_\tau) +
   \sigma_B \right) \mathcal{L}_i\,,
\end{equation}
where $\varepsilon$, $m_\tau$ and $\sigma_B$ are the free parameters
of the fit defined in Sec.~\ref{seq:SCAN},
and $r_i$ is the relative efficiency variation obtained
with the Monte Carlo simulation. The cross section $\sigma(W,m_\tau)$
was calculated according to Eq.~\ref{eq:SigmaTT} with the additional
term describing $\psi'$ production and decay;
it contains $\Gamma_{ee}\!\cdot\!B_{\tau\tau}(\psi')$ as an additional
free parameter.

\begin{table}[t]
  \caption{\label{tab:Points}The summary of the $\tau^{+}\tau^{-}$ threshold
    scan data: $\left<E\right>$, $\delta_E$ -- the time average of
    the beam energy and the corresponding standard deviation,
    $\mathcal{L}$ -- the integrated luminosity,
    $N_{\tau\tau}$ -- the number of events, $\sigma^{obs}_{\tau\tau}$ --
    the observed cross section.
    \vspace*{4pt}
  }
  \newcommand{\NS}{\hspace*{-10pt}}
  \newcommand{\nS}{\hspace*{-6pt}}
  \begin{tabular}{lccrrr}
    \hline\hline\\*[-9pt]
    scan & \NS$\left<E\right>$ & \nS $\delta_E$ & $\mathcal{L}$\phantom{99}
    &\nS $N_{\tau\tau}$
    & $\sigma^{obs}_{\tau\tau}$\phantom{99} \\
    point & \NS (MeV)
    & \nS(MeV) & \hspace*{-6pt}(nb$^{-1}$) &  & (pb)\phantom{xx}\\
\hline\\*[-10pt]
1 &\NS 1771.945 &\nS 0.160 &\NS  668 &\NS  0 & \nS$0.0^{+2.8}$ \\*[2pt]
2 &\NS 1776.408 &\nS 0.086 &\NS 1382 &\NS  1 & \nS$0.7^{+1.7}_{-0.6}$ \\*[2pt]
3 &\NS 1776.896 &\nS 0.045 &\NS 1605 &\NS  6 & \nS$3.7^{+2.2}_{-1.5}$ \\*[2pt]
4 &\NS 1777.419 &\nS 0.061 &\NS 1288 &\NS  4 & \nS$3.1^{+2.5}_{-1.5}$ \\*[2pt]
5 &\NS 1782.103 &\nS 0.060 &\NS  283 &\NS  4 &\nS$14.1^{+11.3}_{-6.8}$ \\*[2pt]
6 &\NS 1792.457 &\nS 0.102 &\NS  233 &\NS  3 &\nS$12.9^{+12.5}_{-7.1}$ \\*[2pt]
7 &\NS 1837.994 &\nS 0.092 &\NS  305 &\NS 14 &\nS$45.8^{+16.0}_{-12.2}$ \\*[2pt]
8($\psi'$) &\NS 1843.040 &\NS 0.065 &\NS 807 &\NS 79 &\nS$97.9^{+11.0}_{-11.0}$\\*[2pt]
9 &\NS 1888.521 &\nS 0.228 &\NS  967 &\NS 49 &\nS$50.7^{+7.2}_{-7.2}$ \\*[4pt]
\hline
\multicolumn{3}{l}{{\it total (excluding $\psi'$)}}        &\NS 6731 &\NS 81 &  \\
    \hline\hline
  \end{tabular}
\end{table}

\vspace*{3pt}
 The fit yielded in
\vspace*{-0.15cm}
\begin{equation}
\begin{split}
 & m_\tau = 1776.80^{+0.25}_{-0.23} \text{ MeV}, \\
 & \varepsilon = 2.29 \pm 0.25 \text{ \%}, \\
 & \sigma_B = 0^{+0.57} \text{ pb}, \\
 & \Gamma_{ee}\!\cdot\!B_{\tau\tau}(\psi') = 7.2 \pm 2.1 \text{ eV,}
\end{split}
\end{equation}

\vspace*{-0.15cm}
\noindent
the background is consistent with zero.

The preliminary estimates of the systematic uncertainties in $m_\tau$
are summarized in Table~\ref{tab:SystErr}. The detector-related
uncertainties, currently dominating, can be substantially reduced with further
data analysis.
\begin{table}[t]
\caption{\label{tab:SystErr}
The preliminary estimates of the systematic uncertainties in the $\tau$ lepton
mass (keV).
\vspace*{4pt}
}
\newcommand{\NS}{\hspace*{-4pt}}
\newcommand{\nS}{\hspace*{-4pt}}
\begin{tabular}{lr}
\hline\hline
\nS Beam energy determination&\NS                40 \\
\nS Detection efficiency variations&\NS         100 \\
\nS Energy spread determination accuracy&\NS     25 \\
\nS Energy dependence of the background&\NS 20 \\
\nS Luminosity measurement instability&\NS       90 \\
\nS Beam energy spread variation&\NS             15 \\
\nS Cross section calculation (r.c., interference)&\NS  30\\
\hline
\nS {\it Sum in quadrature} &\NS 150 \\
\hline\hline
\end{tabular}
\end{table}

\section{Conclusion}

The $\tau$--threshold experiment with the precise beam
energy monitoring is in progress at the VEPP-4M collider with the
KEDR detector.
The preliminary result on the $\tau$ lepton mass
\vspace*{-0.15cm}
$$ m_\tau = 1776.80^{+0.25}_{-0.23} \pm 0.15 \text{ MeV} $$

\vspace*{-0.15cm}
\noindent
is in  good agreement with the world average 
\vspace*{-0.15cm}
$$ m_\tau = 1776.99^{+0.29}_{-0.26} \text{ MeV \cite{PDG2006}} $$

\vspace*{-0.15cm}
\noindent
and has approximately the same accuracy.

Using 0.8~pb$^{-1}$ at the $\psi'$ peak the following preliminary
result was obtained for the $\psi'\to\tau\tau$ decay probability:
\vspace*{-0.3cm}
$$\Gamma_{ee}\!\cdot\!B_{\tau\tau}(\psi') = 7.2 \pm 2.1 \text{ eV,}$$

\vspace*{-0.2cm}
\noindent
The product of the world average values \cite{PDG2006} is
\vspace*{-0.15cm}
$$\left<\Gamma_{ee}\right>\!\cdot\!
\left<B_{\tau\tau}\right>(\psi') = 6.9 \pm 1.7 \text{ eV.}$$

Data taking for this experiment is continued with a goal to
achieve a 0.15~MeV accuracy in the $\tau$ mass. The accuracy of
$\psi'\to\tau\tau$
decay probability will be also well improved.

\end{document}